\documentclass[%
 reprint,
superscriptaddress,
 amsmath,amssymb,
 aps,
pre,
]{revtex4-1}

\usepackage{graphicx}
\usepackage{mathtools} 
\usepackage{siunitx} 

\DeclarePairedDelimiter\floor{\lfloor}{\rfloor}

\newcommand{\modulo}[2]{\modfunction(#1,#2)}
\renewcommand{\modulo}[2]{#1\bmod #2}

\newcommand*\diff{\mathop{}\!\mathrm{d}}

\usepackage{dcolumn}
\usepackage{bm}

\begin{document}


\title{Quantifying fluctuations in reversible enzymatic cycles and clocks}

\author{Harmen Wierenga}
 \email{h.wierenga@amolf.nl}
 \affiliation{AMOLF, Science Park 104, 1098 XG Amsterdam, The Netherlands}
\author{Pieter Rein ten Wolde}
 \affiliation{AMOLF, Science Park 104, 1098 XG Amsterdam, The Netherlands}
\author{Nils B. Becker}
 \affiliation{DKFZ, Bioquant, Im Neuenheimer Feld 267, 69120 Heidelberg, 
Germany}

\date{\today}

\begin{abstract}
Biochemical reactions are fundamentally noisy at a molecular scale. This limits 
the precision of reaction networks, but also allows fluctuation measurements 
which may reveal the structure and dynamics of the underlying biochemical 
network. Here, we study non-equilibrium reaction cycles, such as the 
mechanochemical cycle of molecular motors, the phosphorylation cycle of 
circadian clock proteins, or the transition state cycle of enzymes. Fluctuations 
in such cycles may be measured using either of two classical definitions of the 
randomness parameter, which we show to be equivalent in general microscopically 
reversible cycles. We define a stochastic period for reversible cycles and 
present analytical solutions for its moments. Furthermore, we associate the two 
forms of the randomness parameter with the thermodynamic uncertainty relation, 
which sets limits on the timing precision of the cycle in terms of thermodynamic 
quantities. Our results should prove useful also for the study of temporal 
fluctuations in more general networks.

\end{abstract}

\pacs{Valid PACS appear here}
\maketitle


\section{\label{sec:introduction} Introduction}

Networks of biochemical reactions are at the heart of all biological processes. 
Yet, reactions are stochastic by nature. On the one hand, this poses fundamental 
limits on the precision of any biochemical mechanism. By understanding these 
physical constraints, one can get insights into the design
logic, biological function, and selective pressures that shape the physiology of
living cells. 
On the other hand, stochasticity can be observed experimentally in living cells,
through the advent of a range of experimental techniques which achieve
single-molecule resolution. Fluctuation data complements measurements of
average concentrations, and can aid in model inference~(see 
e.g.\cite{cagatay09}).
Both in finding physical constraints and in performing model inference, we face 
the challenge of devising a theoretical framework for predictions
of fluctuations in biochemical networks.

Here, we present a theoretical study of fluctuations in driven biochemical 
reaction
cycles. Cyclic reactions occur in diverse contexts, such as enzymatic kinetics
with an arbitrary number of intermediates~\cite{Mof14}, molecular motors like 
kinesin~\cite{Svo94}, and molecular clocks
such as the protein modification cycle at the heart of the circadian rhythms of
the cyanobacterium \textit{S. Elongatus}~\cite{Kondo2007}.

Fluctuations in cyclic processes are conventionally quantified by the so-called
randomness parameter, the central quantity of statistical
kinetics~\cite{Svo94,Sch95,Mof14}. The randomness is most easily introduced in
the context of irreversible molecular motors. Here, a motor protein performs a 
sequence of internal rate limiting transitions before each physical step on its 
track. If the last of those transitions is effectively irreversible, then the 
motor can only step forward.
To measure fluctuations in the
stepping rate, the randomness parameter is typically defined as the squared 
coefficient of variation of the waiting time $T$ at each physical position of 
the motor,
\begin{equation}
	r_{T} = \frac{\mathrm{Var \!} \left(T\right)}{\langle T \rangle^{2}}.
	\label{eq:randomness_first_definition}
\end{equation}
Alternatively, we may consider fluctuations in the distance traveled by the
motor in a set time interval. Here one is led to define the randomness as the
Fano factor of the total number of steps $W \! \left(t\right)$ after a (long)
time $t$,
\begin{equation}
	r_{W} = \lim_{t\rightarrow \infty} 
	\frac{\mathrm{Var \! } \left(W \! \left(t\right) \right)}
	{\langle W \! \left(t\right) \rangle}.
	\label{eq:randomness_second_definition}
\end{equation}
It was shown by Schnitzer and Block~\cite{Sch95,Smi53} that in fact these
definitions agree when the cycle is irreversible and $W \! \left(t\right)$
cannot decrease, that is,
\begin{equation}
	\label{eq:rt_equals_rw}
	r_T = r_W.
\end{equation}
In contrast, for reversible cycles, where forward and backward steps in $W \! 
\left(t\right)$ happen with different waiting times, only $r_{W}$ has been 
studied~\cite{Che08}. Its original definition remains valid as the Fano factor 
of the current cycle number. However, it is not obvious suitably extend the 
definition of $r_{T}$,
and whether Eq.\,\ref{eq:rt_equals_rw} can hold. 

In this article, we propose a definition for a stochastic period in reversible
cycles. We show that it possesses properties expected of a period, such as
independence from the initial condition, and equality between the average period 
and
reciprocal of the flux. The randomness parameter $r_T$ associated with this 
period is indeed
equivalent to a step-based randomness $r_W$, so that a generalized
Eq.\,\ref{eq:rt_equals_rw} holds for reversible cycles. This is our first main
result. Next, we give exact expressions for the variance of the period in
general Markovian cycles. Finally, the equivalence Eq.\,\ref{eq:rt_equals_rw}
extends the validity of a recent result, termed the thermodynamic uncertainty 
relation~\cite{Bar15}, to the period of a cycle. Specifically, the relation sets 
a fundamental bound on the precision of the clock period in terms of free energy 
dissipation and substep number. We close by formulating design principles that 
result from this bound, and that show how parameters can be optimized to 
maximize the precision of the period.

\begin{figure*}
	\includegraphics{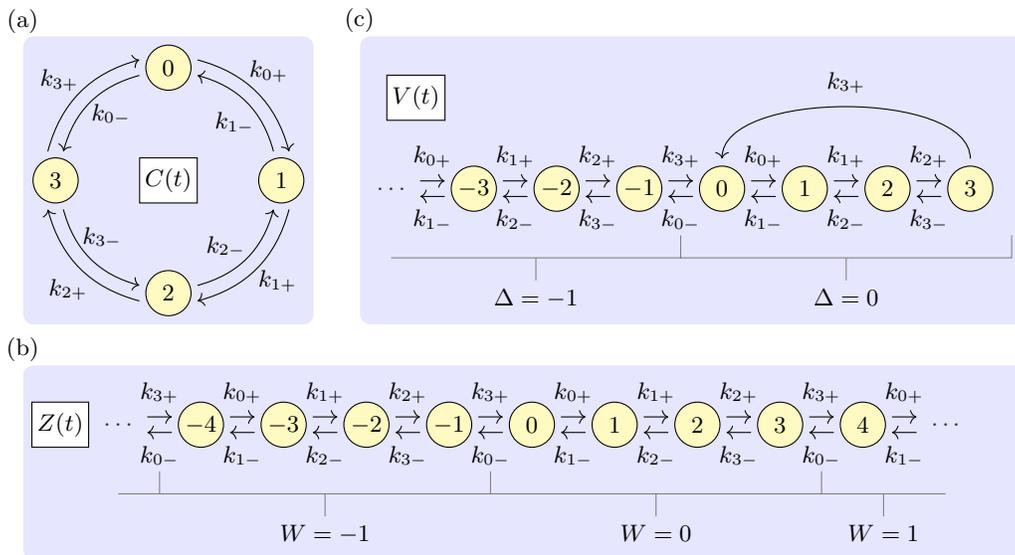}
	\caption{\label{fig:markov_chains}  State
		transition diagrams for three related Markov chains with $N=4$.  
(a) The
		circular reversible Markov chain $C\!\left(t\right)$ with state 
space
		$\left\{0,1,\ldots, N-1\right\}$ and transition rates 
$k_{i\pm}$. We assume there is a drift in the clockwise direction. The average 
period can be
		defined as the reciprocal of the steady state flux. (b) The 
circular chain
		can be unfolded onto the integer line, resulting in the Markov 
chain $Z\!
		\left(t\right) \in \mathbb{Z}$ with initial state $Z \! 
\left(0\right)=0$.
		The stochastic period corresponds to the fist passage time at 
state $N$,
		$T_{0,N}$, which has a well defined probability distribution 
because of the forward drift. The winding
		number $W\! \left(t\right)$ counts the net number of completed 
cycles of the particle. (c) To study the backward fluctuations, it is helpful to 
consider the current
		position $Z\!\left(t\right)$ minus the maximum value of the 
winding number so far, defined as
		the stochastic variable $V\! \left(t\right) \in 
\mathbb{Z}_{<N}$. The
		maximum winding number can be increased by one with rate 
$k_{N-1+}$ from
		state $N-1$, causing $V\! \left(t\right)$ to be reset to $0$. 
The winding
		number relative to its maximum value is given by $\Delta\! 
\left(t\right)$.
	}
\end{figure*}
\section{\label{sec:randomness_parameter} Randomness parameters for reversible
cycles}
Consider a cyclic Markov chain $C\!\left(t\right)$ as in
Fig.~\ref{fig:markov_chains}a. The $N$ states of the Markov process
$C\!\left(t\right)$ are labeled $\left\{0,1,\ldots, N-1 \right\}$, and
possible transitions from state $i$ to state $j$ are defined using the rates
$k_{i,j}$,
\begin{equation}
	k_{i,j}=
	\begin{cases}
		k_{i+} &\text{if } j= \modulo{i+1}{N} \\*
		k_{i-} &\text{if } j= \modulo{i-1}{N} \\*
		-\left(k_{i+}+k_{i-}\right) &\text{if } j=i \\*
		0 &\text{otherwise,}
	\end{cases}
	\label{eq:transition_rates_C}
\end{equation}
where $\modulo{m}{n} \equiv m-n\floor*{\frac{m}{n}}$ is the modulo operation,
and $\floor*{x}$ represents the floor function. The initial location of the
particle is at point $0$, such that $C(0) = 0$. The Markov chain $C(t)$ can be
used to model diverse phenomena, such as the chemical state network of
molecular motors~\cite{Svo94}, or the phosphorylation cycle of the circadian
clock protein KaiC~\cite{Zon07,Pai17}.

Biochemical cycles are driven out of equilibrium by their coupling to an 
external bath. For
example, enzymatic cycles operate out of equilibrium to form products, and are
driven by a replenishment of the substrate or the consumed ATP\@. Here, we
treat fuel turnover implicitly: the use of fuel molecules is absorbed into the
(effective) state transition rates, assuming that the chemical potentials of
fuel molecules are held constant. The total fuel turnover within a cycle
causes a finite free energy drop $\Delta F>0$ per cycle (the affinity of
the cycle) and makes the process macroscopically irreversible, so that
particles proceed in one preferred direction on average. We make the arbitrary
choice that this drift is directed forward. The exponentiated affinity $Q$ is 
related to the transition rates in the cycle by local detailed 
balance~\cite{Tas13},
\begin{equation}
	Q \equiv \exp(-\Delta F/k_BT) = \prod_{i=0}^{N-1} \frac{k_{i-}}{k_{i+}} 
< 1,
	\label{eq:definition_Q}
\end{equation}
where $k_{B}$ and $T$ are the Boltzmann constant and the absolute
temperature, respectively. 

Because the cycle is reversible, the net number of completed cycles, or winding
number, may decrease occasionally. However, the winding number is guaranteed to
ultimately reach any point in the forward direction as a consequence of the
drift (see Sec.\,S.I~\cite{Sup}). Since the particle will move forward, we 
propose to define the stochastic period of the cycle as the time it takes to
move one \emph{net} cycle forward. This implies that succeeding a possible 
backward
cycle, a particle needs to complete two forward cycles to increase the net
position by one and to end the current period. To characterize the period more
clearly we unfold the circular chain onto the integer line as in
Fig.~\ref{fig:markov_chains}b. The Markov chain $Z(t)$ that emerges has state
space $\mathbb{Z}$ and follows similar transition rates as $C(t)$,
\begin{equation}
	\label{eq:transition_rates_Z}
	k_{i,j}=
	\begin{cases}
		k_{(\modulo{i}{N})+} &\text{if } j=i+1 \\*
		k_{(\modulo{i}{N})-} &\text{if } j=i-1 \\*
		-k_{(\modulo{i}{N})+} + k_{(\modulo{i}{N})-} &\text{if } j=i \\*
			0 &\text{otherwise.}
		\end{cases}
\end{equation}
The initial condition reads $Z\!\left(0\right)=0$.

The chain $Z(T)$ contains information not only about the position within the
cycle but also the net number of cycles completed, the winding number $W(t)$,
\begin{equation}
	W\! \left(t\right) = \floor*{\frac{Z \! \left(t\right)}{N}}.
	\label{eq:winding_number}
\end{equation}
The original chain $C\! \left(t\right)$ can be recovered using
the modulo operation,
\begin{align}
	C\! \left(t\right) &= \modulo {Z(t)}{N}
	\nonumber \\*
	& = Z \! \left(t\right) - N W\! \left(t\right).
	\label{eq:recover_C}
\end{align}

We can now consider stochastic waiting times $T_{i,j}$ as first passage times
to reach state $j$ from state $i \neq j$,
\begin{equation}
	\label{eq:perioddef}
	T_{i,j} \equiv \min_{t\geq0}\left\{ t : Z(t) =  j, Z(0)=i \right\}.
\end{equation}
In particular, we define the \emph{stochastic period} of the cycle as
$T_{N} \equiv T_{0,N}$. Backward excursions of arbitrary length are allowed 
within one
period, and need to be reverted completely before a first passage can be made
at point $N$. It can be shown that because of the cyclic symmetry of the chain,
the period is invariant under spatial translations (see Sec.\,S.II~\cite{Sup}),
\begin{equation}
	T_{j,j+N} \sim T_{0,N} \qquad \forall j \in \mathbb{Z},
	\label{eq:period_invariant}
\end{equation}
where the symbol $\sim$ indicates that random variables are identically
distributed. Equation~\ref{eq:period_invariant} shows that the period is an
intrinsic property of the cycle, independent of the starting point, and it 
justifies the shorter notation $T_N$.
The average period is related to the steady state flux $J$ of the circular
Markov chain $C \! \left(t\right)$~\cite{Rei99} (also see 
Sec.\,S.III~\cite{Sup}),
\begin{equation}
\langle T_{N} \rangle = 1/J.
\label{eq:flux_and_period}
\end{equation}

We can now study the randomness parameter in the same manner as was previously
done for irreversible cycles,
\begin{equation}
	r_{T} = \frac{\mathrm{Var \! } \left(T_{N}\right)}{ \langle T_{N} 
\rangle^{2}}.
	\label{eq:randomness_definition_period}
\end{equation}
This definition of the randomness parameter incorporates the original 
Eq.\,\ref{eq:randomness_first_definition}~\cite{Sch95}, and extends it to
reversible systems. For irreversible cycles, the winding number 
$W\!\left(t\right)$ can only increase, and upon an increment of $W\! 
\left(t\right)$, both the first passage time $T_{N}$ and the waiting time $T$ 
end. Therefore, the stochastic period coincides with the waiting time of the 
winding number in this special case.

\begin{figure}
	\includegraphics{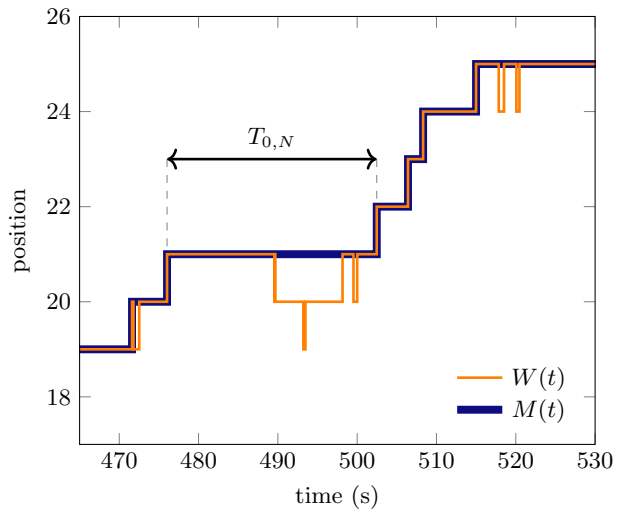}
	\caption{\label{fig:time_trace} 
		Example time traces of the current cycle number of the particle 
(winding number
		$W\! \left(t\right)$), and the furthest position so far (maximum 
$M\!
		\left(t\right)$). The data are produced by kinetic Monte Carlo 
simulations of a
		system with $N=5$ states per cycle, and uniform rates, such that 
on average one
		forward substep is made per \SI{1.0}{\second}, and one backward 
substep per
		\SI{1.5}{\second}. As indicated in the main text, the stochastic 
period $T_{N}$ corresponds to the interval between two increments of $M\! 
\left(t\right)$, and has an average of \SI{15}{\second} in this example.
	}
\end{figure}

The alternative characterization of the fluctuations of the cycle starts from
the winding number. In order to track forward progress only, we consider the
maximal winding number so far,
\begin{equation}
M\! \left(t\right) \equiv \max_{t'\leq t} \left\{ W\! \left(t'\right) \right\}.
\label{eq:definition_M}
\end{equation}
For a molecular motor, the maximal winding number represents the furthest
physical location reached yet. Because it can only increase, and subsequent
values are independent and identically distributed, $M\! \left(t\right)$ is a
renewal process~\cite{Smi53}. As shown in Fig.\,\ref{fig:time_trace}, intervals 
of constant $M$ end whenever the
winding number reaches a value that was not attained before. The stochastic
waiting time of $M\! \left(t\right)$, defined as the time between two
increments, is precisely $T_{N}$, because a first passage at the next cycle ends 
the period and raises the maximum winding number. Accordingly, $M\! 
\left(t\right)$ also counts the number of completed periods so far.

We may then define a randomness based on the maximal winding number as
\begin{equation}
	\label{eq:randomness_definition_M}
	r_M = \lim_{t\to\infty}
	\frac{\mathrm{Var} \! \left(M\! \left(t\right)\right)}%
	{\langle M\! \left(t\right) \rangle}.
\end{equation}
In the special case of irreversible cycles, $r_M$ as defined here reduces to
the previous winding number randomness $r_W$
Eq.\,\ref{eq:randomness_second_definition}, since then $M\to W$. For general
reversible cycles, $r_W$ and $r_M$ are not a priori equal.

\section{\label{sec:results} Results}

\subsection{\label{sec:randomness_proof} Equivalence of randomness parameters}

Schnitzer and Block showed that $r_W$ and $r_T$ coincide for irreversible
cycles, Eq.\,\ref{eq:rt_equals_rw}~\cite{Sch95}. 
It turns out that their proof carries over unchanged to the reversible
system, when using the maximal winding number instead of the winding
number. That is,
\begin{equation}
	r_T = r_M.
\label{eq:randomness_equivalence_M}
\end{equation}
The original proof of this relation was given by Smith~\cite{Smi53} in the
context of general renewal processes. It holds because $M\! \left(t\right)$ is
a renewal process with waiting times $\sim T_{N}$, as indicated before. In the 
following, we will show that in fact
not only Eq.\,\ref{eq:randomness_equivalence_M} but also
Eq.\,\ref{eq:rt_equals_rw} holds for reversible cycles.

To that end, we first need to prove the following relation,
\begin{equation}
\lim_{t\rightarrow \infty} 
\frac{\mathrm{Var} \! \left(W\! \left(t\right)\right)}{\langle W\! 
\left(t\right) \rangle} 
= \lim_{t\rightarrow \infty} \frac{\mathrm{Var} \! 
\left(M\! \left(t\right)\right)}{\langle M\! \left(t\right) \rangle}.
\label{eq:equivalence_fano_factors}
\end{equation}
It is convenient to study the nonpositive difference $\Delta\! \left(t\right) 
\equiv W\!
\left(t\right) - M\! \left(t\right)$, which we use to rewrite
\begin{align}
r_W &=\lim_{t\rightarrow \infty} 
\! \frac{\mathrm{Var}\!\left[W\! \left(t\right)\right]}{\langle W\! 
\left(t\right) \rangle} \nonumber \\* 
&= \lim_{t\rightarrow \infty} \! \frac{\mathrm{Var}\!\left[M\! 
\left(t\right)\right] + \mathrm{Var}\!\left[\Delta\! \left(t\right)\right]+ 2 
\mathrm{Cov}\!\left[M\! \left(t\right), \Delta\! \left(t\right)\right]}{\langle 
M\! \left(t\right) \rangle + \langle \Delta\! \left(t\right) \rangle}.
\label{eq:expansion_Fano_factor}
\end{align}
Both the mean and variance of $M\! \left(t\right)$ grow linearly with $t$ in
leading order~\cite{Smi53}. We will show below that the mean and variance of 
$\Delta\!
\left(t\right)$ stay finite for long times, implying that they do not
contribute to the limit, Eq.\,\ref{eq:expansion_Fano_factor}. The remaining
covariance term has bounds set by the Cauchy-Schwarz inequality~\cite{Wil91},
\begin{equation}
\lvert \mathrm{Cov}\!\left[M\! \left(t\right), \Delta\! \left(t\right)\right] 
\rvert \leq \sqrt{\mathrm{Var}\!\left[M\! \left(t\right)\right] 
\mathrm{Var}\!\left[\Delta\! \left(t\right)\right]}.
\label{eq:cauchy-schwarz}
\end{equation}
If the variance of $\Delta \! \left(t\right)$ remains finite, then the 
covariance
term in Eq.\,\ref{eq:expansion_Fano_factor} grows slower than the leading
term  $\mathrm{Var}\!\left[M\! \left(t\right)\right]$, and will not contribute
to the limit either. Then indeed Eq.\,\ref{eq:randomness_equivalence_M}
follows.

Hence, it remains to show that the first two statistical moments of $\Delta \!
\left(t\right)$ actually stay finite in the long time limit. To this end we
introduce the new microscopic variable $V \! \left(t\right)$, defined as
\begin{equation}
V \! \left(t\right) = Z \! \left(t\right) - N M \! \left(t\right).
\label{eq:definition_V}
\end{equation}
$V \! \left(t\right)<N$ is the current offset from the furthest cycle reached. 
Our motivation for studying $V \! \left(t\right)$ is its relation to $\Delta \! 
\left(t\right)$,
\begin{align}
\floor*{ \frac{V \! \left(t\right)}{N} } &= \floor*{ \frac{Z \! 
\left(t\right)}{N} } - M \! \left(t\right)  \nonumber \\*
&= W \! \left(t\right) - M \! \left(t\right) \nonumber \\*
&= \Delta \! \left(t\right).
\label{eq:relation_V_D}
\end{align}
Since $M \! \left(t\right)$ is derived from $Z \! \left(t\right)$, the new
variable $V \! \left(t\right)$ is completely determined by $Z \! 
\left(t\right)$. In fact, by itself $V \! \left(t\right)$ constitutes a Markov 
chain, with a state transition diagram as shown in 
Fig.~\ref{fig:markov_chains}c. As shown there, one transition of this chain is 
irreversible.

To understand the one-way transition, we notice that the maximum winding number 
increases irreversibly from $M$ to $M+1$ whenever $Z \! \left(t\right)$ 
transitions from state $\left(M+1\right)  N-1$ to state
$\left(M+1\right) N$. Concurrently, $V \! \left(t\right)$ transitions from
state $N-1$ to state $0$ (Eq.\,\ref{eq:definition_V}), effectively resetting it
to the origin. Therefore, the irreversibility of the transition originates from 
that of $M \! \left(t\right)$.

Using the new Markov chain, we can show that the mean and variance of $V \! 
\left(t\right)$ and $\Delta \! \left(t\right)$ will tend to finite values. In 
contrast to $Z \! \left(t\right)$ on the integer line, which increases endlessly 
and thus has no stationary distribution, $V \! \left(t\right)$ does have a 
stationary distribution. As shown in detail
in Sec.\,S.IV~\cite{Sup}, the stationary
probability distribution decreases exponentially for $V \! \left(t\right) \to 
-\infty $. Then, Eq.\,\ref{eq:relation_V_D} implies that $\Delta  \! 
\left(t\right)$ also has a
stationary distribution with an exponential tail on the left. In other words,
it is exponentially unlikely for the particle to move backwards a certain 
distance from the furthest point reached before. Because the stationary 
distribution of $\Delta  \! \left(t\right)$ is bounded on the
right and has an exponential tail on the left, it has finite moments of any
order. In particular, this implies that as $t\to\infty$, the mean and variance
of $\Delta  \! \left(t\right)$ are $\mathcal{O}\left(t^{0}\right)$, so that the
covariance term in Eq.\,\ref{eq:expansion_Fano_factor} is
$\mathcal{O}\left(t^{1/2}\right)$. This completes the proof of
Eq.\,\ref{eq:equivalence_fano_factors}, and consequently establishes relation
Eq.\,\ref{eq:randomness_equivalence_M}, so that in conclusion, we arrive at 
Eq.\,\ref{eq:rt_equals_rw},
\begin{equation}
	\label{eq:allrequal}
	r \equiv r_T = r_M = r_W
\end{equation}
for general reversible or irreversible reaction cycles.


\subsection{\label{sec:analytical_comparison} Statistical moments of the
period}

The randomness parameter is useful for comparing experimental results to 
theoretical predictions. Chemla et al.~derived the exact solution $r_W$ for 
reversible cycles~\cite{Che08}, and this exact
solution now carries over to $r_T$. Additionally, their analytical expression 
provides an alternative route for showing the equivalence of both forms of the 
randomness parameter. In this section we give analytical solutions of the 
moments of the first passage time,
which directly give $r_T$. Comparing this result with that for $r_W$, we
confirm the equivalence Eq.\,\ref{eq:allrequal}.

To find the moments of $T_{N}$, it is convenient to decompose the first
passage time into first passage times at each microscopic interval,
\begin{equation}
	T_{N} \sim T_{0,N} = \sum_{i=0}^{N-1} T_{i,i+1}.
	\label{eq:decomposition_FPT}
\end{equation}
Notice that a path causing a first passage is allowed to recross the initial
point an arbitrary number of times before reaching the final point. We denote
the first-passage time density of $T_{i,j}$ by $f_{i,j}\!
\left(t\right)$. $S_{i}\! \left(t\right)$ denotes the survival probability at
state $i$, given by \(S_i(t) = \exp(-(k_{i+}+k_{i-})t)\). The single step first
passage time distributions are linked by the relation
\begin{equation}
	f_{i,i+1}\! \left(t\right) = 
	S_{i}\! \left(t\right) k_{i+} \! + \! \int_{0}^{t} \! \diff t' S_{i} 
\left(t'\right)
	k_{i-} f_{i-1, i+1}\! \left(t-t'\right).
	\label{eq:fp_dist_defining_eq}
\end{equation}
Eq.\,\ref{eq:fp_dist_defining_eq} splits the paths from $i$ to $i+1$ into paths
where the first step is made in either the positive or negative direction. By 
taking the Laplace transform and using it as the moment generating function, we 
find relations for all statistical moments of $T_{i,i+1}$. In 
Sec.\,S.V~\cite{Sup},
these coupled equations are derived and solved, providing exact solutions to
the mean and variance of the stochastic period,
\begin{align}
	\langle T_{N} \rangle 
	&=  \left(1-Q\right)^{-1} \sum_{i,j=0}^{N-1} \frac{1}{k_{i+}} 
	\prod_{\ell=1}^{j} 
\frac{k_{\left(i+\ell\right)-}}{k_{\left(i+\ell\right)+}};  \nonumber \\
	\mathrm{Var} \! \left(T_{N}\right) 
	&= \left(1-Q\right)^{-3} \sum_{i=0}^{N-1} 
	\left[ \sum_{j=0}^{N-1} \frac{1}{k_{\left(i-j\right)+}} 
\prod_{\ell=1}^{j} \frac{k_{\left(i-j+\ell\right)-}}{k_{\left(i-j+\ell\right)+}} 
\right]^{2}
	\nonumber \\ 
	& \qquad \times \left(1+2\sum_{j=1}^{N-1} \prod_{\ell=1}^{j} 
\frac{k_{\left(i+\ell\right)-}}{k_{\left(i+\ell\right)+}} + 
		\prod_{\ell=0}^{N-1} \frac{k_{\ell-}}{k_{\ell+}}\right).
\label{eq:general_moments}
\end{align}
Here, $Q$ is the product of the rates as in Eq.\,\ref{eq:definition_Q}.

We inserted the analytical moments Eq.\,\ref{eq:general_moments} in
Eq.\,\ref{eq:randomness_definition_period} to obtain an exact expression for
$r_T$. We then compared this to the analytical expression for $r_{W}$ obtained 
by Chemla et al.~\cite{Che08}. While Eq.\,\ref{eq:allrequal} implies that they 
must be identical, we were unable to rewrite our analytical expression in the 
form of that of Chemla et al. due to the complexity of the expressions. Still, 
using the computer algebra system \emph{Mathematica}~\cite{Mathematica}, we have 
directly confirmed their equivalence up to $N=55$, suggesting it holds for 
arbitrary $N$ (as Eq.\,\ref{eq:allrequal} implies), and lending strong support 
for Eq.\,\ref{eq:general_moments} and the result by Chemla et al.


\subsection{\label{sec:optimization_randomness} Optimal precision of the period}

\begin{figure}
	\includegraphics{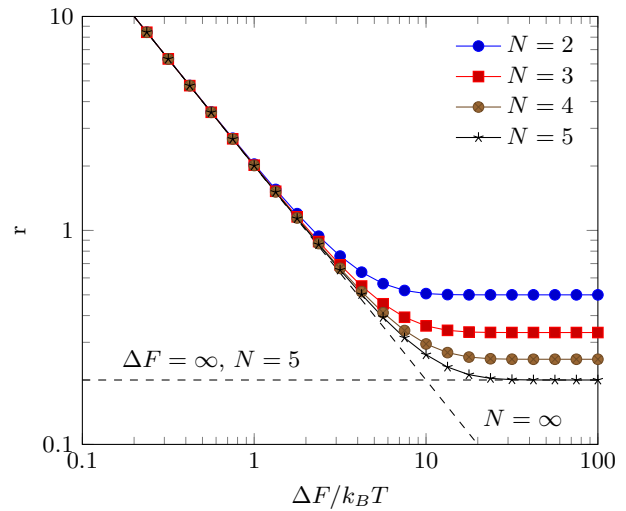}
	\caption{\label{fig:varianceVsDissipation} Dissipation suppresses
		fluctuations. The randomness parameter is calculated in the 
optimal case of
		uniform rates, as a function of the number of steps per cycle 
$N$ and the
		dissipation per cycle $\Delta F$. It is evident that for low
		dissipation, increasing the number of steps cannot lower the 
randomness
		parameter. However, when the free energy drop per cycle becomes 
large ($\Delta F \geq 2 N k_{B} T$), $N$ becomes limiting and the precision 
increases only with increasing the number of substeps. The dotted lines 
represent the final inequality in Eq.\,\ref{eq:uncertainty_relation} for the 
case $N=5$.}
\end{figure}
In recent work, Barato and Seifert introduced a so-called
thermodynamic uncertainty relation, which bounds the minimum value of the Fano
factor $r_{W}$ in the long time limit~\cite{Bar15} in terms of the number of 
substates
and the energy dissipation rate. The bound was subsequently proved
by Gingrich et al.~\cite{Gin16}. Combined with our result 
Eq.\,\ref{eq:allrequal}, the bound reads
\begin{align}
	r_T &= r_W
	\nonumber \\* 
	&\geq \frac{1}{N} \coth \left(\frac{\Delta F}{2N k_{B} T} \right) 
\nonumber \\* 
	&\geq \max \left\{\frac{2 k_{B} T}{\Delta F},\frac{1}{N}\right\}.
	\label{eq:uncertainty_relation}
\end{align}
The first inequality is the thermodynamic uncertainty 
relation~\cite{Bar15,Gin16}; it is saturated for a cycle with uniform forward 
and
backward rates, such that $k_{i\pm}=k_{\pm}$. The second, looser inequalities
are found in the limit that either $N$ or $\Delta F$ approaches infinity. Hence, 
the equivalence of the different randomness parameters Eq.\,\ref{eq:allrequal} 
extends the uncertainty relation to the cycle period, which therefore has direct 
implications for the workings of molecular clocks.

The bound Eq.\,\ref{eq:uncertainty_relation} for the precision of timing $r_T$ 
constrains the optimal design of biochemical timekeeping
mechanisms. Circadian clocks are present in many organisms, such as the
relatively simple cyanobacteria \cite{Gro86}. Here, a
post-translational modification cycle produces approximately $24$ hour rhythms
\cite{Nak05}. Although in real cyanobacteria many KaiABC protein complexes in a
bacterium go through their biochemical modification cycles in a partly
synchronized way, we can expect that the single-cycle expression
Eq.\,\ref{eq:uncertainty_relation} gives a correct intuition about the limiting
resources: A precise clock period requires both a sufficient number of cyclic
substates and an adequate cycle affinity. When both the average period of the
cycle and the free energy dissipation per cycle are held constant, then the
lowest possible variance is achieved by choosing uniform rates 
\cite{Bar15,Gin16}. Furthermore, to prevent wasting resources, it is beneficial 
to tune both the number of steps
$N$ and the free energy dissipation, such that both bounds on the final line of 
Eq.\,\ref{eq:uncertainty_relation} are equally limiting,
\begin{equation}
	N_{\mathrm{opt}}=\frac{\Delta F_{\mathrm{opt}}}{2 k_{B} T}.
	\label{eq:optimal_resource_allocation}
\end{equation}
Fig.~\ref{fig:varianceVsDissipation} shows that when $\Delta F/k_{B}T$ is small, 
it limits the precision, meaning that the optimal randomness is independent of 
$N$. In other words, the low free energy investment cannot be compensated by a 
large number of substeps. Conversely, when $\Delta F/k_{B}T$ is larger than 
$2N$, it has hardly any benefit in reducing the randomness. The number of 
substeps then limits the precision, and the latter can only increase by 
increasing $N$. A similar design principle has recently been obtained for the 
optimal allocation of resources in cellular sensing~\cite{Gov14a,Gov14b}. In an 
optimally designed system, each fundamental resource should be equally limiting.

\subsection{\label{sec:measurement_protocols} Measurement protocols}

The two equivalent forms $r_T = r_W$ of the randomness parameter relate to two
alternative methods of measuring the stochasticity of an enzymatic 
cycle~\cite{Sch95}. A prolonged measurement of the time trace of a
single motor protein position may include many ($M$) completed periods, giving 
$M$
samples for computing the randomness parameter $r_T$ (see 
Fig.\,\ref{fig:time_trace}). However, as pointed out by Schnitzer and 
Block~\cite{Sch95}, experimental measurements of each period may require a high 
temporal and spatial resolution. In the example of a motor protein, it would be 
essential to track the timing and direction of each physical step. When such 
measurement precision is not available, $r_{W}$ gives another option for 
measuring the randomness, since it only requires samples of the number of 
completed cycles $W \! \left(t\right)$. For a motor, this corresponds to the 
final distance traveled after a certain time, while for enzymatic cycles it 
coincides with the number of product molecules. Multiple independent samples are 
needed in order to measure $r_{W}$ via 
Eq.\,\ref{eq:randomness_second_definition}. This data can either be gained from 
a collection of propagating molecules, or from sections of a single long time 
trace. This way, when the resolution to measure individual periods and to 
estimate $r_{T}$ is missing, $r_{W}$ can still provide an estimate for the 
randomness parameter.

Now, considering the case where the measurement precision is sufficient for 
determining either $r_{T}$ or $r_{W}$, is there one that converges quicker to 
the desired result? In Sec.\,S.VI~\cite{Sup} we compare the estimation of 
$r_{W}$ and $r_{T}$ from similar sets of data, to investigate the differences 
between their estimation errors. We use a simple method to calculate both 
versions of the randomness parameter, and show that estimation of $r_{W}$ 
involves a trade-off in the finite observation time $t$ after which samples of 
$W\!\left(t\right)$ are taken. For short $t$, many samples of 
$W\!\left(t\right)$ are found, but the result suffers from a significant 
systematic bias. For large $t$, the bias disappears since we approach the limit 
in definition Eq.\,\ref{eq:randomness_second_definition}, but the number of 
samples decreases, reducing the precision. The estimate of $r_{T}$ is not 
subject to this trade-off, and appears to converge quicker than that of $r_{W}$, 
because $T_{N}$ can be sampled more frequently than $W\!\left(t\right)$. 
Therefore, $r_{T}$ should be favored when the experimental resolution allows it; 
use of $r_{W}$ requires careful correction of the finite-time bias.

\section{\label{sec:discussion} Discussion}

In recent works, efforts were made to study temporal precision in
stochastic networks. Barato and Seifert considered either the Fano factor 
$r_W$~\cite{Bar15c, Bar16}, or the coherence time of oscillations in the 
occupancy of a single
site~\cite{Bar17}. Here we have presented the equivalence 
Eq.\,\ref{eq:allrequal} for reversible cycles, showing that the Fano factor 
$r_W$ can be
reinterpreted as a clock precision by relating it to the fluctuations in the 
period $r_{T}$. This in turn widens the applicability of the thermodynamic 
uncertainty
relation~\cite{Bar15} to temporal variables. In the example of molecular clocks, 
it translates to a bound for the clock precision of reversible cycles. A cycle 
reaches the highest timing accuracy when all transitions have uniform forward- 
and backward rates. As
intuitively expected, the bound shows that the period can become more 
deterministic
when either the number of substates per cycle or the free energy dissipation per 
cycle
is increased. Yet both these quantities are fundamentally limiting the
precision, meaning that each sets its own lower bound on the precision, and
that one cannot compensate for the other~\cite{Gov14a,Gov14b}. 

The second study~\cite{Bar17} employs a different approach
to quantifying the precision of biochemical clocks, which can be applied to any 
non-equilibrium network, and involves defining an
average period as a property of the transition matrix. That average period is
not to be confused with the stochastic period $T_{N}$ defined here, which is
a random variable; their relation in the case of cyclic networks is an 
interesting open question.

In a different work, Barato and Seifert discuss the information contained in 
higher moments of the period and propose bounds on the skewness and kurtosis for 
irreversible cycles~\cite{Bar15b}. We show how to calculate these moments for 
reversible cycles in section~\ref{sec:analytical_comparison}. It would be 
interesting to see whether tighter lower bounds exist for the higher moments 
just as for the randomness parameter, and whether those bounds are saturated for 
a system with uniform rates.

Very recently, Gingrich and Horowitz reported a relation between the
large deviation functions for currents and first passage times in
general Markov chains~\cite{Gin17}, which applies to the present setup. They 
also made a connection between the thermodynamic uncertainty relation and first 
passage time statistics. The main contribution of our present work is a proof of 
the identification $r_{T} = r_{W}$, Eq.\,\ref{eq:allrequal} for unicyclic 
networks. Using the concept of a stochastic period, our direct derivation of 
this equivalence also affords physical intuition for the relation of current and 
first-passage time in cycles.

It is intriguing to ask if the equivalence Eq.\,\ref{eq:allrequal} holds for
more general network topologies than the reversible cycle as shown in
Fig.\,\ref{fig:markov_chains}a. Indeed, by examining the derivations one
can see that Eq.\,\ref{eq:allrequal} holds in more general networks if they
fulfill the following conditions. First, it must be possible to define a
winding number $W \! \left(t\right)$, of which the average has a forward drift. 
This condition is already fulfilled by taking $W \! \left(t\right)$ as the net 
number of times a
specific transition in the network is made, whenever there is a stationary net
forward flux through this pathway. The renewal process $M \! \left(t\right)$
derived from the winding number then obeys
Eq.\,\ref{eq:randomness_equivalence_M}, where the period is defined as the
waiting time of the renewal process. Second, the difference $\Delta \!
\left(t\right)$ between the current- and maximum position should have a limiting
distribution with a finite average and variance. This condition ensures that
Eq.\,\ref{eq:equivalence_fano_factors} holds. It remains to be seen which
networks conform to these conditions, and whether the first condition implies
the second.

\begin{acknowledgments}
We thank Giulia Malaguti for careful reading of the manuscript. 
This work is part of the research programme of the Netherlands Organisation for 
Scientific Research (NWO) and was performed at the research institute AMOLF.
\end{acknowledgments}

%


\widetext
\pagebreak
\begin{center}
\textbf{\large Supplemental Material: \\ 
	Quantifying fluctuations in reversible enzymatic cycles and clocks}
\end{center}

\setcounter{equation}{0}
\setcounter{figure}{0}
\setcounter{table}{0}
\setcounter{page}{1}
\setcounter{section}{0}
\makeatletter
\renewcommand{\theequation}{S.\arabic{equation}}
\renewcommand{\thesection}{S.\Roman{section}}
\renewcommand{\thefigure}{S.\arabic{figure}}
\renewcommand{\bibnumfmt}[1]{[S#1]}
\renewcommand{\citenumfont}[1]{S#1]}


\section{\label{sec:direction} Proof that the particle moves forward}

The cyclic Markov chain $C\!\left(t\right)$ and its unwound version 
$Z\!\left(t\right)$, as defined in Eq.\,4 and Eq.\,6 of the main text 
respectively, have several interesting properties. In this section, we will show 
that the non-equilibrium driving of the cycle, which is set by the exponentiated 
affinity $Q<1$, logically implies that a particle will make an arbitrary number 
of forward transitions with probability $1$. This fact is used for the 
definition of the period in Sec.\,II of the main text.

As mentioned in the main text, $T_{i,j}$ is the first passage time at state $j$ 
given that the particle started in state $i$, and $f_{i,j}\!\left(t\right)$ is 
its probability density function. We denominate the probability that a particle 
will make at least one passage at $j$ by $\alpha_{i,j}$. This fraction is given 
by the integral of the first passage time density over all time, and thus the 
goal of this section is to show that
\begin{equation}
\alpha_{i,j} \equiv \int_{0}^{\infty} \! \diff t f_{i,j} \! \left(t\right) = 1 
\qquad \mathrm{if} \; i<j \; \mathrm{and} \; Q<1.
\label{eq:definition_alpha}
\end{equation}
Before we arrive at this result, we describe some of the characteristics of the 
Markov chain $Z\!\left(t\right) \in \mathbb{Z}$. 

First, we would like to find relations for the first passage time distributions 
$f_{i,j}\!\left(t\right)$, with $i<j$. For this, it is helpful to define the 
survival probability at a specific point as
\begin{equation}
S_{i} \! \left(t\right) = e^{-\left(k_{i+}+k_{i-}\right)t}.
\label{eq:survival_probability} 
\end{equation}
The survival probability is defined as the probability that a particle remains 
at location $i$ for at least a time $t$, meaning that it has not transitioned 
yet. When this probability is multiplied with the rate of making transitions, 
one finds the probability density of waiting times at state $i$. The first step, 
which ends such a waiting time, can either be in the forward or in the backward 
direction. Accordingly, the first passage time density from $i$ to $j$ can be 
conditioned on the direction of this first step. If the step moves the particle 
from $i$ to $i\pm1$ at exactly time $t'$, then the particle still needs to 
travel from state $i\pm1$ to $j$ in precisely the remaining time $t-t'$, in 
order to complete the first passage. This provides the following fundamental 
relation on the single step first passage times,
\begin{equation}
f_{i,j} \! \left(t \right) = \int_{0}^{t} \diff t' \left[S_{i} \! \left(t' 
\right) k_{i+} f_{i+1, j} \! \left(t-t'\right) +  S_{i} \! \left(t' \right) 
k_{i-} f_{i-1, j} \! \left(t-t'\right) \right].
\label{eq:decomposition_arbitrary_steps}
\end{equation}
For $i=j-1$ the equation changes, because a single forward step creates a first 
passage directly,
\begin{equation}
f_{j-1,j} \! \left(t \right) = S_{j-1} \! \left(t \right) k_{j-1+} + 
\int_{0}^{t} \diff t' S_{j-1} \! \left(t' \right) k_{j-1-} f_{j-2, j} \! 
\left(t-t'\right).
\label{eq:decomposition_single_step}
\end{equation}
These equations can be used to show that the particle will always move forward. 
The following proof of Eq.\,\ref{eq:definition_alpha} resembles that given by 
e.g. Anderson~\cite{And11}. 

A condition for the $\alpha_{i,j}$ is found by integrating both sides of 
Eq.\,\ref{eq:decomposition_arbitrary_steps} over all $t\geq0$. This integral 
turns the convolutions into simple products of the separate integrals, and the 
integral of Eq.\,\ref{eq:survival_probability} can be easily computed, leading 
to
\begin{equation}
\alpha_{i,j} = \frac{k_{i+}}{k_{i+}+k_{i-}} \alpha_{i+1,j} + 
\frac{k_{i-}}{k_{i+}+k_{i-}} \alpha_{i-1,j},
\label{eq:alpha_1}
\end{equation}
whereas for $i=j-1$, we find
\begin{equation}
\alpha_{j-1,j} = \frac{k_{j-1+}}{k_{j-1+}+k_{j-1-}} + 
\frac{k_{j-1-}}{k_{j-1+}+k_{j-1-}} \alpha_{j-2,j}.
\label{eq:alpha_2}
\end{equation}
The denominators of Eq.\,\ref{eq:alpha_1} and Eq.\,\ref{eq:alpha_2} can be 
brought to their respective left hand sides, revealing that
\begin{align}
\left(\alpha_{i,j}-\alpha_{i-1,j} \right) &= \frac{k_{i+}}{k_{i-}} 
\left(\alpha_{i+1,j}-\alpha_{i,j}\right) \qquad \mathrm{if} \> i<j-1 \nonumber 
\\
\left(\alpha_{j-1,j}-\alpha_{j-2,j}\right) &= \frac{k_{j-1+}}{k_{j-1-}} 
\left(1-\alpha_{j-1,j} \right).
\label{eq:alpha_3}
\end{align}
Recursive use of this equation gives
\begin{equation}
\left(\alpha_{i,j}-\alpha_{i-1,j} \right) = \left(1-\alpha_{j-1,j} \right) 
\prod_{\ell=i}^{j-1} \frac{k_{\ell+}}{k_{\ell-}}.
\label{eq:alpha_4}
\end{equation}
This allows us to express each probability in terms of $\alpha_{j-1,j}$,
\begin{align}
\alpha_{i,j} &= \alpha_{j-1,j} + \sum_{h=i+1}^{j-1} \left(\alpha_{h-1,j} - 
\alpha_{h,j} \right) \nonumber \\
&= \alpha_{j-1,j} - \left(1-\alpha_{j-1,j} \right) \sum_{h=i+1}^{j-1} 
\prod_{\ell=h}^{j-1} \frac{k_{\ell+}}{k_{\ell-}},
\label{eq:alpha_5}
\end{align}
where we have to remember that we are only considering the case that $i<j$. The 
formula can be checked for $i=j-1$, in which case the summation from $j$ to 
$j-1$ vanishes by definition of the sum, and only the first term survives. 

Next, it is possible to take the limit of $i \rightarrow -\infty$. Furthermore, 
notice that
\begin{equation}
\sum_{h=-\infty}^{j-1} x_{h} = \sum_{m=-\infty}^{0} \sum_{n=j-N}^{j-1} x_{m N+n}
\label{eq:modulo_decomposition_sum}
\end{equation}
for any $x_{h}$, and 
\begin{equation}
\prod_{\ell = mN+n}^{j-1} x_{\ell} = \left(\prod_{\ell = mN+n}^{n-1} 
x_{\ell}\right) \left(\prod_{\ell=n}^{j-1} x_{\ell} \right) = \left(\prod_{\ell 
= mN}^{-1} x_{\ell}\right) \left(\prod_{\ell=n}^{j-1} x_{\ell} \right),
\label{eq:splitting_product}
\end{equation}
where the last equality is true when $x_{h+mN}=x_{h}$. This step becomes clear 
by rearranging the front and the back of the product to make it independent of 
$n$. Eq.\,\ref{eq:modulo_decomposition_sum} and Eq.\,\ref{eq:splitting_product} 
imply that
\begin{align}
\lim_{i \rightarrow -\infty} \alpha_{i,j} &= \alpha_{j-1,j} - 
\left(1-\alpha_{j-1,j} \right) \sum_{m=-\infty}^{0} \sum_{n=j-N}^{j-1} 
\prod_{\ell=m N + n}^{j-1} \frac{k_{\ell+}}{k_{\ell-}} \nonumber \\
&= \alpha_{j-1,j} - \left(1-\alpha_{j-1,j} \right) \Bigg[ \sum_{n=j-N}^{j-1} 
\prod_{\ell=n}^{j-1} \frac{k_{\ell+}}{k_{\ell-}} \Bigg] \Bigg[ 
\sum_{m=-\infty}^{0} \prod_{\ell=m N}^{-1} \frac{k_{\ell+}}{k_{\ell-}} \Bigg] 
\nonumber \\
&= \alpha_{j-1,j} - \left(1-\alpha_{j-1,j} \right) \Bigg[ \sum_{n=j-N}^{j-1} 
\prod_{\ell=n}^{j-1} \frac{k_{\ell+}}{k_{\ell-}} \Bigg] \sum_{m=0}^{\infty} 
\left(Q^{-1}\right)^{m}.
\label{eq:alpha_6}
\end{align}
In the final step, we used the definition of $Q$ from Eq.\,5 of the main text to 
rewrite the product over $N$ subsequent fractions of rates. Additionally, the 
cyclic symmetry of the rates $k_{m N+l\pm} = k_{l\pm}$ was applied to see that 
$Q$ emerges at each interval. The final sum of Eq.\,\ref{eq:alpha_6} diverges, 
since we made the choice $Q<1$. However, the limit of $\alpha_{i,j}$ must be a 
number between $0$ and $1$, because it represents a probability. The only way 
this is possible is when $\alpha_{j-1,j}=1$, because this takes out the 
diverging term before the limit is taken. Now the proof is easily finalized, as 
Eq.\,\ref{eq:alpha_5} shows that
\begin{equation}
\alpha_{i,j} = 1 \qquad \forall \> i<j.
\label{eq:alpha_7}
\end{equation}
%


\section{\label{sec:period_invariant} Period invariance under translations} 

To ensure a proper definition of the period in Sec.\,II of the main text, it is 
important that the cycle completion time is not influenced by the initial 
position. If a different period existed for each initial state, this would 
disagree with the intuition that oscillations are a property of the system as a 
whole. In the following, we show that the stochastic time interval $T_{0,N}$ 
complies with this requirement.

Firstly, the transition rates in the system are symmetric under translations by 
multiples of $N$, causing the statistics of the first passage times to obtain an 
identical symmetry,
\begin{equation}
T_{i+mN,j+mN} \sim T_{i,j} \qquad \forall m \in \mathbb{Z}.
\label{eq:translation_symmetry_system}
\end{equation}
Here, $\sim$ means that both variables are identically distributed.  Focusing on 
$T_{0,N}$, a first passage at $N>1$ requires passages at all intermediate 
points. Hence, the stochastic period can be decomposed into the first passage 
times between all unique forward transitions,
\begin{equation}
T_{0,N} = \sum_{i=0}^{N-1} T_{i,i+1}.
\label{eq:decomposition_period}
\end{equation}
Furthermore, the strong Markov property implies that all intermediate first 
passage times are statistically independent, showing that we can solve the 
distribution of the full period by calculating the distributions of the 
intermediate intervals. This will be important in 
Sec.\,\ref{sec:analytical_solutions}, where we calculate the statistical moments 
of the period.

To show that the period is independent from the initial state, the general 
initial position is taken to be any $j \in \mathbb{Z}$. 
Eq.\,\ref{eq:translation_symmetry_system} implies that this initial position can 
be chosen such that $0 \leq j \leq N-1$ without influencing the distribution. 
Then, the first passage time is decomposed as in 
Eq.\,\ref{eq:decomposition_period},
\begin{align}
T_{j,j+N} &= \sum_{i=j}^{j+N-1} T_{i,i+1} \nonumber \\*
&= \sum_{i=j}^{N-1} T_{i,i+1} + \sum_{i=N}^{j+N-1} T_{i,i+1} \nonumber \\*
&\sim \sum_{i=j}^{N-1} T_{i,i+1} + \sum_{i=0}^{j-1} T_{i,i+1} \nonumber \\*
&= \sum_{i=0}^{N-1} T_{i,i+1} = T_{0,N}. \label{eq:decomposition_period_initial}
\end{align}
In this notation, a summation from $0$ to $-1$ vanishes. In going from the 
second to the third line, Eq.\,\ref{eq:translation_symmetry_system} is used to 
lower the index by $N$. As a result, it is shown that the initial position $j$ 
has no influence on the time it takes to complete a net full cycle. Hence, the 
period in unambiguously defined, and it is justified to call it $T_{N} \equiv 
T_{0,N} \sim T_{j,j+N}$, without indicating the initial position.


\section{\label{sec:flux} Connection between flux and average period}

The average period $\langle T_{0,N} \rangle$ can be related to the steady state 
flux of the cycle $C\!\left(t\right)$. To show this connection, we follow a 
strategy that is similar to the one showing the equivalence between $r_{T}$ and 
$r_{W}$ in the main text. Since the maximum winding number $M\!\left(t\right)$ 
is a renewal process, we can apply the elementary renewal theorem, which 
states~\cite{Smi53}
\begin{equation}
\lim_{t\rightarrow \infty} \frac{\langle M \! \left(t\right) \rangle}{t} = 
\lim_{t\rightarrow \infty} \frac{d \langle M \! \left(t\right) \rangle}{d t} = 
\frac{1}{\langle T_{0,N} \rangle}.
\label{eq:elementary_renewal_theorem}
\end{equation}
The first step is due to l'H\^{o}pital's rule. A connection can be made with the 
current winding number $W \! \left(t\right)$, using the definition of $\Delta \! 
\left(t\right)$,
\begin{equation}
\lim_{t\rightarrow \infty} \frac{d \langle W \! \left(t\right) \rangle}{d t} = 
\lim_{t\rightarrow \infty} \frac{\langle W \! \left(t\right) \rangle}{t} = 
\lim_{t\rightarrow \infty} \frac{\langle M \! \left(t\right) \rangle + \langle 
\Delta \! \left(t\right) \rangle}{t} = \lim_{t\rightarrow \infty} \frac{\langle 
M \! \left(t\right) \rangle}{t},
\label{eq:average_winding_numbers}
\end{equation}
which uses the fact that the average of $\Delta \! \left(t\right)$ becomes 
constant in steady state, as shown in Sec.\,\ref{sec:stationary distribution}, 
while the average of $M \! \left(t\right)$ increases linearly with time in 
leading order. The latter follows from the elementary renewal theorem, see e.g. 
Smith~\cite{Smi53}. The left hand side of the equation can be expressed in terms 
of $Z \! \left(t\right)$ and its probability distribution $p_{i} \! 
\left(t\right) \equiv \mathbb{P}\left\{Z\!\left(t\right)=i\right\}$,
\begin{align}
\frac{d \langle W \! \left(t\right) \rangle}{d t} &= \frac{d}{d t} 
\sum_{m=-\infty}^{\infty} m \sum_{i=0}^{N-1} p_{mN+i} \! \left(t\right) 
\nonumber \\
&= \sum_{m=-\infty}^{\infty} m \sum_{i=0}^{N-1} \left[ 
-\left(k_{i+}+k_{i-}\right)p_{mN+i} \! \left(t\right)+k_{i-1+}p_{mN+i-1} \! 
\left(t\right)+k_{i+1-}p_{mN+i+1} \! \left(t\right) \right] \nonumber \\
&= \sum_{m=-\infty}^{\infty} m \left[-k_{N-1+}p_{\left(m+1\right)N-1} \! 
\left(t\right) -k_{0-} p_{mN} \! \left(t\right) +k_{N-1+} p_{mN-1} \! 
\left(t\right) +k_{0-} p_{\left(m+1\right)N} \! \left(t\right) \right] \nonumber 
\\
&= \sum_{m=-\infty}^{\infty} \left[m - \left(m-1\right)\right] \left[k_{N-1+} 
p_{mN-1} \! \left(t\right) -k_{0-} p_{mN} \! \left(t\right) \right] \nonumber \\
&= k_{N-1+} \sum_{m=-\infty}^{\infty} p_{mN-1} \! \left(t\right) -k_{0-} 
\sum_{m=-\infty}^{\infty} p_{mN} \! \left(t\right).
\label{eq:av_W_decomposition}
\end{align}
In the second line, we use the transition matrix from Eq.\,6 of the main text to 
evaluate the time derivative. Then, the terms in the sum over $i$ that cancel 
each other are taken out. Finally, we relabel $m$ to $m-1$ in the terms where 
the factor $m+1$ is present in the label, such that the final result emerges. 
The two sums over probabilities can be reinterpreted in connection with the 
Markov process on the circular state space. Eq.\,8 of the main text implies that 
the probability that $C \! \left(t\right)$ equals $i$ can be expanded as a sum 
over the possible values of $Z \! \left(t\right)$,
\begin{equation}
\mathrm{P} \left\{ C \! \left(t\right) = i\right\} = \sum_{m=-\infty}^{\infty} 
p_{mN+i} \! \left(t\right).
\label{eq:C_sum_of_Z}
\end{equation}
Since $C \! \left(t\right)$ has a finite state space, it has a limiting 
distribution, which we call $\pi_{i}$. Combining all previous equations, and 
taking the long time limit, we find a connection between the steady state flux 
through the cycle $J$, and the average first passage time,
\begin{equation}
\frac{1}{\langle T_{0,N} \rangle} = k_{N-1+} \pi_{N-1} - k_{0-} \pi_{0} = J.
\label{eq:flux_time_connection}
\end{equation}
We make use of this fact in Sec.\,II of the main text. Notice that $J$ is 
constant through the cycle by definition of the steady state, i.e.
\begin{equation}
J = k_{i+} \pi_{i} - k_{i+1-} \pi_{i+1} \quad \forall i.
\end{equation}
Combined with Eq.\,\ref{eq:flux_time_connection}, this also implies 
Eq.\,\ref{eq:decomposition_period_initial}. Hence, we may redefine the cycle 
start point arbitrarily.


\section{\label{sec:stationary distribution} Stationary distribution of 
$V\!\left(t\right)$}

In this section, we show that the Markov process $V\!\left(t\right)$, which is 
shown in Fig.\,1 of the main text, has a stationary distribution. Furthermore, 
the exact form of the stationary probability will be derived, showing that it 
has an exponentially decaying tail. The latter property will be used to show 
that the moments of $\Delta\!\left(t\right)$ will remain constant in the long 
time limit, which is an important step in Sec.\,III\,A of the main text.

First, we need to show that the stationary distribution exists. For this, we use 
the definition of the steady state flux
\begin{equation}
J_{i} \equiv k_{i+} \phi_{i} - k_{i+1-} \phi_{i+1}
\label{eq:flux_definition}
\end{equation}
for $i<N-1$, where $\phi_{i}$ is the sought after stationary distribution of 
$V\!\left(t\right)$. Additionally, a special definition is used for the only 
irreversible step of the chain,
\begin{equation}
J_{N-1} \equiv k_{N-1+} \phi_{N-1}.
\label{eq:flux_definition_end}
\end{equation}
The stationary distribution is the solution of the following set of equations,
\begin{align}
J_{i}-J_{i-1} = 0 \qquad &\mathrm{for} \quad i < 0 \nonumber \\
J_{0}-J_{-1}-J_{N-1} = 0 \qquad &\mathrm{for} \quad i=0 \nonumber \\
J_{i}-J_{i-1} = 0 \qquad &\mathrm{for} \quad 0 < i \leq N-1 \nonumber \\
\sum_{i=-\infty}^{N-1} \phi_{i} = 1. \qquad&
\label{eq:stationary_distribution_condition}
\end{align}
The fist equation shows that
\begin{equation}
J_{i-1} = J_{i} = J_{-1} \qquad \forall i<0.
\label{eq:left_flux}
\end{equation}
The second equality follows from applying an inductive step to the first 
equality. Hence, the flux in the infinite chain is constant. Additionally, the 
stationary distribution only exists if the final normalization condition of 
Eq.\,\ref{eq:stationary_distribution_condition} is obeyed. For this, it is 
required that the stationary distribution vanishes infinitely far away,
\begin{equation}
\lim_{i\rightarrow -\infty} \phi_{i} = 0.
\label{eq:limit_stationary_distribution}
\end{equation}
Together with Eq.\,\ref{eq:flux_definition} and Eq.\,\ref{eq:left_flux}, it 
implies that
\begin{equation}
J_{i} = 0 \qquad \forall i<0.
\label{eq:detailed_balance_chain}
\end{equation}
Thus, if the stationary distribution exists, then detailed balance is obeyed on 
the negative integer tail. In this case, the flux equations in 
Eq.\,\ref{eq:stationary_distribution_condition} reduce to those of a cycle 
without the integer tail,
\begin{align}
J_{0}-J_{N-1} = 0 \qquad &\mathrm{for} \quad i=0 \nonumber \\
J_{i}-J_{i-1} = 0 \qquad &\mathrm{for} \quad 0 < i \leq N-1.
\label{eq:stationary_distribution_condition_circle}
\end{align}
Consequently, the part of the solution for $i\geq0$ is proportional to the 
stationary probability of a cycle with an equal transition matrix. Previously, 
exact solutions of this stationary distribution were reported for arbitrary 
rates, which were derived using various methods~\cite{Der83,Ada97}. Therefore, 
by inserting the rates including the irreversible step, 
Eq.\,\ref{eq:stationary_distribution_condition_circle} can be solved in terms of 
$\phi_{0}$. In addition, Eq.\,\ref{eq:detailed_balance_chain} for the tail 
probabilities has a simple solution in terms of $\phi_{0}$,
\begin{equation}
\phi_{i} = \phi_{0} \prod_{\ell=i}^{-1} \frac{k_{\ell+1-}}{k_{\ell+}},
\label{eq:solution_stationary_tail}
\end{equation}
which holds for $i\leq0$. Both solutions, for the circle and the tail, are 
connected at $\phi_{0}$. We can combine them by normalization, to obtain the 
full solution of Eq.\,\ref{eq:stationary_distribution_condition},
\begin{equation}
\phi_{i} = \begin{cases}
\frac{1}{\mathcal{N}_{\phi}} \left(\sum_{j=0}^{N-1} \prod_{\ell=1}^{j} 
\frac{k_{\ell-}}{k_{\ell+}} \right) \prod_{m=0}^{-\left(i+1\right)} 
\frac{k_{-m-}}{k_{-\left(m+1\right)+}} \text{ for } i \leq 0 \\
\frac{1}{\mathcal{N}_{\phi}} \frac{k_{0+}}{k_{i+}} \left(\sum_{j=0}^{N-i-1} 
\prod_{\ell=1}^{j} \frac{k_{i+\ell-}}{k_{i+\ell+}} \right) \text{ for } 0 \leq i 
\leq N-1.
\end{cases}
\label{eq:solutions_stationary_distribution} 
\end{equation}
Notice that the two definitions coincide for $i=0$. These equations solve all 
linear equations Eq.\,\ref{eq:stationary_distribution_condition}, with as many 
variables as conditions, and thus provide a unique solution. However, it was 
assumed that a solution exists, and this is only true if the normalization 
constant converges,
\begin{align}
\mathcal{N}_{\phi} &= \sum_{i=1}^{N-1} \frac{k_{0+}}{k_{i+}} 
\left(\sum_{j=0}^{N-i-1} \prod_{\ell=1}^{j} \frac{k_{i+\ell-}}{k_{i+\ell+}} 
\right) \nonumber \\
&+ \sum_{i=-\infty}^{0} \left(\sum_{j=0}^{N-1} \prod_{\ell=1}^{j} 
\frac{k_{\ell-}}{k_{\ell+}} \right) \prod_{m=0}^{-\left(i+1\right)} 
\frac{k_{-m-}}{k_{-\left(m+1\right)+}}.
\label{eq:normalisation_stationary_distribution}
\end{align}
The first sum has finite bounds, so it trivially converges. The second sum 
consists of a factor times the following series,
\begin{align}
&\sum_{i=-\infty}^{0} \prod_{m=0}^{-\left(i+1 \right)} 
\frac{k_{-m-}}{k_{-\left(m+1\right)+}} \nonumber \\ 
&= \sum_{i=0}^{\infty} \prod_{m=0}^{i-1} \frac{k_{-m-}}{k_{-\left(m+1\right)+}} 
\nonumber \\
&= \sum_{j=0}^{\infty} \sum_{i=0}^{N-1} \prod_{m=0}^{i+jN-1} 
\frac{k_{-m-}}{k_{-\left(m+1\right)+}} \nonumber \\
&= \sum_{j=0}^{\infty} \sum_{i=0}^{N-1} \prod_{m=0}^{jN-1} 
\frac{k_{-m-}}{k_{-\left(m+1\right)+}} \prod_{n=jN}^{jN+i-1} 
\frac{k_{-n-}}{k_{-\left(n+1\right)+}} \nonumber \\
&= \left( \sum_{i=0}^{N-1} \prod_{n=0}^{i-1} 
\frac{k_{-n-}}{k_{-\left(n+1\right)+}} \right) \sum_{j=0}^{\infty}  
\prod_{m=0}^{jN-1} \frac{k_{-m-}}{k_{-\left(m+1\right)+}} \nonumber \\
&= \left(\sum_{i=0}^{N-1} \prod_{n=0}^{i-1} 
\frac{k_{-n-}}{k_{-\left(n+1\right)+}}\right) \sum_{j=0}^{\infty}  Q^{j}.
\label{eq:infinite_series_convergence}
\end{align}
On the third line, the sum over $i$ is decomposed into a sum over the cycles, 
and a sum within the cycle. Then, after splitting the product into two parts, 
the fifth line makes use of the periodic property of the rates, causing 
$k_{-n+jN\pm}=k_{-n\pm}$. On the final line, $Q$ is recognized in a slightly 
different form, again valid due to periodicity,
\begin{equation}
Q = \prod_{\ell=0}^{N-1} \frac{k_{\ell-}}{k_{\ell+}} = \prod_{m=0}^{N-1} 
\frac{k_{-m-}}{k_{-\left(m+1\right)+}}.
\label{eq:Q_alternative_form}
\end{equation}
In the final form of Eq.\,\ref{eq:infinite_series_convergence}, the geometric 
series is uncovered. Since the choice $Q<1$ was made in the definition of the 
Markov process $Z\!\left(t\right)$, the series converges. Hence, a unique 
stationary distribution exists, and is given by 
Eq.\,\ref{eq:solutions_stationary_distribution}. Moreover, for negative $i$, it 
can be seen that the limiting probability decays exponentially with the distance 
from $i=0$,
\begin{equation}
\phi_{i-N} = Q \phi_{i}.
\label{eq:exponential_decay_phi}
\end{equation}
This provides the proof that the moments of the difference to the maximum remain 
finite. The probability distribution for $\Delta\!\left(t\right)$ can be 
expressed in terms of the probability distribution for $V\!\left(t\right)$,
\begin{equation}
\mathrm{P}\left\{\Delta\!\left(t\right)=-n\right\} = \sum_{i=0}^{N-1} 
\mathrm{P}\left\{ V\!\left(t\right) = -n N +i\right\}
\label{eq:delta_rel_V}
\end{equation}
for $n \leq 0$. For large $t$, this probability will approach its stationary 
distribution, and Eq.\,\ref{eq:exponential_decay_phi} guarantees that it decays 
exponentially too,
\begin{equation}
\mathrm{P} \left\{\Delta \left(\infty\right)= -n\right\} = Q^{n-1} \mathrm{P} 
\left\{\Delta \left(\infty\right)=-1\right\}.
\label{eq:exponential_decay_Delta}
\end{equation}
Since all moments of any distribution with exponential tails are finite, and 
$V\!\left(t\right)$ reaches its stationary state in the infinite time limit, 
this completes the proof that the mean and variance of $\Delta\!\left(t\right)$ 
that occur in Eq.\,17 of the main text will remain finite.


\section{\label{sec:analytical_solutions} Analytical solutions of the 
statistical moments of the period}

For many applications, from mathematical proofs to numerical evaluation, it is 
helpful to have analytical expressions of key quantities. Here, we study the 
stochasticity of the period, which is generally quantified using the statistical 
moments of its distribution, and provide a method to calculate these moments. 
For example, the mean and variance of the period are derived, which are utilized 
in Sec.\,III\,B of the main text.

From Eq.\,\ref{eq:decomposition_period} and the statistical independence of the 
$T_{j,j+1}$, it follows that the mean and variance of $T_{0,N}$ can be split 
into those of the single step first passage times,
\begin{align}
\langle T_{0,N} \rangle &= \sum_{i=0}^{N-1} \langle T_{i,i+1} \rangle \nonumber 
\\
\mathrm{Var}\left( T_{0,N} \right) &= \sum_{i=0}^{N-1} \mathrm{Var}\left( 
T_{i,i+1} \right).
\label{eq:decomposition_moments} 
\end{align}
The decomposition of the random variable $T_{0,N}$ in 
Eq.\,\ref{eq:decomposition_period}  is equivalent to separating the distribution 
$f_{0,N} \! \left(t\right)$ using an N-fold convolution,
\begin{equation}
f_{0,N} \! \left(t\right) = 	
\int_{0}^{t} \prod_{i=0}^{N}\Bigg[ \diff t_{i} \, f_{i,i+1} \! 
\left(t_{i}\right) \Bigg] \delta \! \left(t - \sum_{i=0}^{N-1} t_{i} \right).
\label{eq:conv_decomp}
\end{equation}
Here, $\delta \! \left(t\right)$ represents the Dirac delta function, and the 
product affects both the variables of integration and the single step first 
passage time densities. The Laplace transform of Eq.\,\ref{eq:conv_decomp} gives
\begin{equation}
\widetilde{f}_{0,N} \! \left(s\right)  = \prod_{i=0}^{N-1}
\widetilde{f}_{i,i+1} \! \left(s\right).
\label{eq:conv_decomp_lap}
\end{equation}
This equation will be useful for finding the closed form expressions for the 
statistical moments of $T_{0,N}$. 

To start the derivation of the moments, we need a relation on the probability 
distribution of the period, for which we recall 
Eq.\,\ref{eq:decomposition_single_step},
\begin{equation}
f_{i,i+1} \! \left(t \right) = S_{i} \! \left(t \right) k_{i+} + \int_{0}^{t} 
\diff t' S_{i} \! \left(t' \right) k_{i-} f_{i-1, i+1} \! \left(t-t'\right).
\label{eq:decomposition_single_step_2}
\end{equation}
To remove the convolution, we apply the Laplace transform to both sides of this 
relation. The final term can be decomposed using the same approach as in 
Eq.\,\ref{eq:conv_decomp_lap},
\begin{equation}
\widetilde{f}_{i-1,i+1} \! \left(s \right)=\widetilde{f}_{i-1,i} \! \left(s 
\right) \widetilde{f}_{i,i+1} \! \left(s \right).
\label{eq:conv_decomp_lap_2}
\end{equation}
This produces a set of $N$ second order equations in $\widetilde{f}_{j j+1} \! 
\left(s \right)$,
\begin{equation}
\left(s+k_{i+}+k_{i-}\right) \widetilde{f}_{i,i+1} \! \left(s\right) = k_{i+} + 
k_{i-} \widetilde{f}_{i-1,i} \! \left(s \right) \widetilde{f}_{i,i+1} \! \left(s 
\right).
\label{eq:decomposition_single_step_laplace}
\end{equation}
Since the equations are of second order, there will be two solutions. The 
appropriate solution can be chosen by realizing that the drift, set by $Q<1$, 
causes the particles to move forward. Sec.\,\ref{sec:direction} shows that the 
choice $Q<1$ ensures that a first passage in the forward direction is made with 
probability 1, and thus $\widetilde{f}_{j,j+1} \! \left(0 \right) = 
\alpha_{j,j+1}=1$. It becomes clear immediately by setting $s=0$ in 
Eq.\,\ref{eq:decomposition_single_step_laplace} that $\widetilde{f}_{j,j+1} \! 
\left(0 \right) =1$ solves the equation at this specific point.

Moments of the first-passage time can be calculated by evaluating derivatives of 
the Laplace transformed distribution at $s=0$. Hence, relations for the moments 
are obtained by applying the derivative operator to 
Eq.\,\ref{eq:decomposition_single_step_laplace}. For $n\geq1$, the $n$th order 
derivative of Eq.\,\ref{eq:decomposition_single_step_laplace} gives the 
following relation,
\begin{equation}
\left(s+k_{i+}+k_{i-} \right) \widetilde{f}_{i,i+1}^{(n)} \! \left(s\right) + n 
\widetilde{f}_{i,i+1}^{(n-1)} \! \left(s\right) = k_{i-} \sum_{\ell=0}^{n} 
\binom{n}{\ell} \widetilde{f}_{i-1,i}^{(\ell)} \! \left(s\right) 
\widetilde{f}_{i,i+1}^{(n-\ell)} \! \left(s \right).
\label{eq:derivatives_relation}
\end{equation}
These equations become relevant by setting $s=0$, using the moment generating 
property of the Laplace transform and $\widetilde{f}_{j,j+1} \! \left(0 \right) 
=1$. Upon a slight reorganisation, relations are uncovered for the moments of 
the single step first passage times,
\begin{equation}
k_{i+}\langle T_{i,i+1}^{n} \rangle - k_{i-} \langle T_{i-1,i}^{n} \rangle = n 
\langle T_{i,i+1}^{n-1} \rangle + k_{i-} \sum_{\ell=1}^{n-1} \binom{n}{\ell} 
\langle T_{i-1,i}^{\ell} \rangle \langle T_{i,i+1}^{n-\ell} \rangle. 
\label{eq:relation_moments}
\end{equation}
Similar relations were found for the moments in birth-death processes on the 
positive integers~\cite{Jou08}. Notice that the unknown averages on the left 
hand side are of order $n$, while those on the right hand side are at most of 
order $n-1$. We can therefore solve this set of equations recursively to obtain 
the moments to any desired order. In the following we demonstrate this for the 
mean and variance. The relation for the average first passage time reads
\begin{equation}
k_{i+}\langle T_{i,i+1} \rangle - k_{i-} \langle T_{i-1,i} \rangle = 1, 
\label{eq:relation_average}
\end{equation}
and for the mean square first passage time
\begin{equation}
k_{i+}\langle T_{i,i+1}^{2} \rangle - k_{i-} \langle T_{i-1,i}^{2} \rangle = 2 
k_{i-} \langle T_{i,i+1} \rangle \langle T_{i-1,i} \rangle +2\langle T_{i,i+1} 
\rangle.
\label{eq:first_relation_variance}
\end{equation}
Using Eq.\,\ref{eq:relation_average}, the latter can be rewritten to a simpler 
form,
\begin{equation}
k_{i+}\langle T_{i,i+1}^{2} \rangle - k_{i-} \langle T_{i-1,i}^{2} \rangle = 2 
k_{i+} \langle T_{i,i+1} \rangle^{2}.
\label{eq:relation_variance}
\end{equation}
The solution for the variance can then be inferred from the solution of the mean 
squared first passage times. By dividing Eq.\,\ref{eq:relation_moments} by 
$k_{i+}$, it becomes clear that the general form of the linear equations fits 
the following simple linear relation, where the variables $X_{i}$ exist for all 
$i\in \left\{0,1,...,N-1\right\}$ and the parameters $\lambda_{i}$ are allowed 
to be different for each $i$,
\begin{equation}
X_{i}-\lambda_{i} X_{i-1} = \beta_{i}.
\label{eq:general_difference}
\end{equation}
The indices are considered to increment modulo $N$. It can be checked by 
substitution that the general solution of this set of $N$ linear relations is 
given by
\begin{equation}
X_{i}=\left[1-\prod_{\ell=0}^{N-1}\lambda_{\ell}\right]^{-1} \sum_{j=0}^{N-1} 
\beta_{i-j} \prod_{\ell=1}^{j} \lambda_{i-j+\ell}.
\label{eq:general_difference_solution}
\end{equation}
This solution is applied to Eq.\,\ref{eq:relation_average} and 
Eq.\,\ref{eq:relation_variance}, which results in
\begin{align}
\langle T_{i,i+1} \rangle &=  \left(1-\prod_{\ell=0}^{N-1} 
\frac{k_{\ell-}}{k_{\ell+}}\right)^{-1} \sum_{j=0}^{N-1} \frac{1}{k_{i-j+}} 
\prod_{\ell=1}^{j} \frac{k_{i-j+\ell-}}{k_{i-j+\ell+}}
\nonumber \\
\langle T_{i,i+1}^{2} \rangle &= \left(1-\prod_{\ell=0}^{N-1} 
\frac{k_{\ell-}}{k_{\ell+}}\right)^{-1} \sum_{j=0}^{N-1} 2 \langle T_{i-j,i-j+1} 
\rangle^{2} \prod_{\ell=1}^{j} \frac{k_{i-j+\ell-}}{k_{i-j+\ell+}}. 
\label{eq:variance_partial_fpt}
\end{align}
Using Eq.\,\ref{eq:decomposition_moments}, the previous equations can be used to 
give the average and variance of the full cycle first-passage time. After 
adjusting the summation variables, the general formulas for the average and 
variance of the period become
\begin{align}
\langle T_{N} \rangle &=  \left(1-\prod_{\ell=0}^{N-1} 
\frac{k_{\ell-}}{k_{\ell+}}\right)^{-1} \sum_{i,j=0}^{N-1} \frac{1}{k_{i+}} 
\prod_{\ell=1}^{j} \frac{k_{i+\ell-}}{k_{i+\ell+}}  \nonumber \\
\mathrm{Var}\left(T_{N}\right) &= \left(1-\prod_{\ell=0}^{N-1} 
\frac{k_{\ell-}}{k_{\ell+}}\right)^{-1} \sum_{i,j=0}^{N-1} 2 \langle T_{i,i+1} 
\rangle^{2} \prod_{\ell=1}^{j} \frac{k_{i+\ell-}}{k_{i+\ell+}} - 
\sum_{i=0}^{N-1} \langle T_{i,i+1} \rangle^{2} \nonumber \\
&= \left(1-\prod_{\ell=0}^{N-1} \frac{k_{\ell-}}{k_{\ell+}}\right)^{-1} 
\sum_{i=0}^{N-1} \langle T_{i,i+1} \rangle^{2} \left(1+2\sum_{j=1}^{N-1} 
\prod_{\ell=1}^{j} \frac{k_{i+\ell-}}{k_{i+\ell+}} + \prod_{\ell=0}^{N-1} 
\frac{k_{\ell-}}{k_{\ell+}}\right) \nonumber \\
&= \left(1-\prod_{\ell=0}^{N-1} \frac{k_{\ell-}}{k_{\ell+}}\right)^{-3} 
\sum_{i=0}^{N-1} \left[ \sum_{j=0}^{N-1} \frac{1}{k_{i-j+}} \prod_{\ell=1}^{j} 
\frac{k_{i-j+\ell-}}{k_{i-j+\ell+}} \right]^{2} \left(1+2\sum_{j=1}^{N-1} 
\prod_{\ell=1}^{j} \frac{k_{i+\ell-}}{k_{i+\ell+}} + \prod_{\ell=0}^{N-1} 
\frac{k_{\ell-}}{k_{\ell+}}\right). \label{eq:general_variance_fpt}
\end{align}
%


\section{\label{sec:statistics_randomness_parameter} Two protocols with 
different statistics for measurement of Randomness Parameter}

\begin{figure}
	\includegraphics{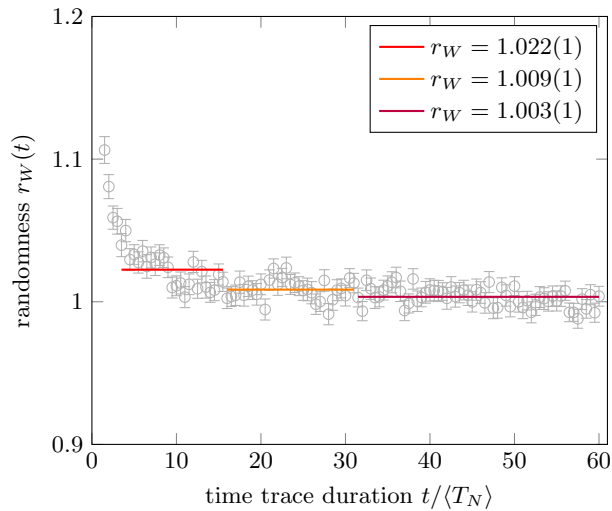}
	\caption{\label{fig:rWvsObservationTime} Approximate Fano factor 
$r_{W}\!\left(t\right)$ as a function of the finite observation time $t$. 
$r_{W}\!\left(t\right)$ is estimated from kinetic Monte Carlo simulations of a 
cycle with uniform rates and parameter values $N=5$, 
$k_{+}=\SI{1.0}{\per\second}$, and $k_{-}=\SI{0.67}{\per\second}$. For this 
system, the average period $\langle T_{N} \rangle = \SI{15}{\second}$, and the 
true randomness parameter $r=1$. We generate $200$ time traces of length $t$, 
with initial condition $W\!\left(0\right)=0$ and $Z\!\left(0\right)$ uniformly 
distributed from $0$ to $N-1$, to correspond to the stationary distribution 
within the cycle. From each trace we sample $W\!\left(t\right)$ and then 
calculate $r_{W}\!\left(t\right)$ via Eq.\,\ref{eq:finite_time_Fano_factor}. 
This whole procedure is repeated $25$ times for each time trace duration to 
gather statistics, after which we report the mean and standard error of the mean 
(SEM) of the estimate $r_{W}\!\left(t\right)$ in gray. Clearly, the Fano factor 
$r_{W}\!\left(t\right)$ contains a systematic bias for small observation times. 
In red, orange and purple, we average over all points in the specified time 
windows, and show in the legend that a small, decreasing bias persists even for 
larger observation times. Values for $r_{W}$ are reported as mean(SEM).}
\end{figure}
In Sec.\,III\,A of the main text, we show the existence of two equivalent 
expressions of the randomness parameter. These two versions also correspond to 
different protocols for measuring the randomness parameter from observations of 
a particle that is moving through a cycle. The first protocol, which estimates 
$r_{T}$, involves measuring the time $T_{N}$ it takes to complete each cycle. 
Hence, enough measurement precision is needed to determine when a cycle is 
completed, but the single substeps within the cycle need not be resolved. In the 
example of a motor protein, this means that each physical step of the motor is 
seen, and the time it takes to reach the furthest point so far is tracked. 
Hence, this protocol needs positional data to be collected at a frequency much 
faster than the average period of one cycle. When collecting samples of $T_{N}$, 
it is important to avoid a selection bias.
That is, one should collect a fixed number of periods, and not all
periods within a fixed total time, which would bias the measurement by
favoring short periods. This bias decreases with total time, but is best
avoided since its magnitude is not known a priori.

The second protocol measures $r_{W}$. Here, samples of the winding number 
$W\!\left(t\right)$ are taken at a certain finite time $t$, since the infinite 
time limit in Eq.\,2 of the main text cannot be achieved experimentally. For 
motor proteins, the distance traveled after a time $t$ is sampled. All 
information on the single cycle times is lost in this picture, requiring more 
particles or long, separable time traces to get enough statistics for 
determining $r_{W}$. Still, a lower resolution is required to perform the second 
protocol, since it only the accumulation of many completed cycles needs to be 
measured. Therefore, the protocol to measure $r_{W}$ is more accessible than 
that of $r_{T}$.

Now, consider a case where enough resolution is available to choose either 
protocol. Are there benefits for choosing one over the other? In the remainder 
of this section, we will investigate the accuracy at which both protocols 
estimate the randomness parameter, and show that $r_{T}$ has clear benefits over 
$r_{W}$.

To calculate the precision of the two protocols, we perform Kinetic Monte Carlo 
simulations of cycles with uniform rates. As explained in 
Fig.\,\ref{fig:rWvsObservationTime}, we choose parameters such that the 
randomness parameter $r=1$, and sampled $r_{W}\!\left(t\right)$ and $r_{T}$. The 
second protocol can only sample at finite times, and we define the Fano factor 
of $W\!\left(t\right)$ as
\begin{equation}
r_{W}\!\left(t\right) = 
\frac{\mathrm{Var}\!\left(W\!\left(t\right)\right)}{\langle 
W\!\left(t\right)\rangle},
\label{eq:finite_time_Fano_factor}
\end{equation}
and the true randomness parameter is given by
\begin{equation}
r_{W} = \lim_{t\rightarrow \infty} r_{W}\!\left(t\right).
\label{eq:randomness_limit}
\end{equation}
First, we investigate the deviations between $r_{W}\!\left(t\right)$ and its 
limit $r_{W}$. In Fig.\,\ref{fig:rWvsObservationTime}, we show how $r_{W}\! 
\left(t\right)$ approaches the asymptote. A systematic bias is clearly visible 
for small observation times, but seems to flatten out after roughly ten periods 
for the chosen parameter set. Averages over several intervals weighted with the 
inverse variance show that a bias of roughly $1\%$ persists even for an 
observation time of $20$ periods.

Now, we can choose to estimate $r_{W}\! \left(t\right)$ at either $10$ or $20$ 
periods, and compare the estimation error to that of $r_{T}$ taken from a 
similar amount of data. We take 300 time traces for each sample of $r_{W}$ of 
length $10\langle T_{N}\rangle$ or $20\langle T_{N}\rangle$. Correspondingly, we 
estimate $r_{T}$ from either $300\times10$ or $300\times20$ samples of $T_{N}$. 
This procedure creates a single sample of $r_{T}$ and of $r_{W}$, and we 
repeated it $300$ times to investigate the error made when only a single sample 
of $r$ would be taken in an experiment. We found the following estimates for $r$ 
and its spread (mean(SEM) $\pm$ standard deviation) for data from time traces of 
$10\langle T_{N}\rangle$,
\begin{align}
r_{W} &= 1.054(4) \pm 0.087 \nonumber \\
r_{T} &= 0.999(2) \pm 0.056.
\label{eq:estimates_r_reversible_10}
\end{align}
Take note that the final numbers are the standard deviations of the 
distribution, not of the mean, which represent the accuracy of a single 
measurement of $r$. It is clear that the measurement of $r_{W}$ is biased, while 
the one of $r_{T}$ is not. The accuracies of the two measurements are 
comparable, with the spread in $r_{W}$ being $1.5$ times larger than that in 
$r_{T}$. For time traces of $20\langle T_{N}\rangle$, the bias on $r_{W}$ 
becomes smaller, but more data is available for $r_{T}$, which makes it more 
precise,
\begin{align}
r_{W} &= 1.029(3) \pm 0.085 \nonumber \\
r_{T} &= 0.997(2) \pm 0.038.
\label{eq:estimates_r_reversible_20}
\end{align}
The spread in $r_{W}$ is now $2.3$ times larger than that in $r_{T}$. This shows 
that there is a trade-off in choosing the measurement time; a short sample time 
allows for many measurements, but there is a systematic bias on the result, 
while a long sample time makes use of only a small part of the available data, 
but has a smaller bias.

We repeated the previous analysis for an irreversible cycle with $N=5$, 
$k_{+}=\SI{0.33}{\per\second}$, and $k_{-}=0$. This system has $\langle T_{N} 
\rangle = \SI{15}{\second}$, and $r=0.2$. This repetition is done to investigate 
the influence of $\Delta\!\left(t\right)$ on the precision, because it vanishes 
for irreversible cycles. We find for a sampling time of $10\langle T_{N}\rangle$ 
that
\begin{align}
r_{W} &= 0.2128(7)  \pm 0.018 \nonumber \\
r_{T} &= 0.2002(2) \pm 0.0056.
\label{eq:estimates_r_irreversible_10}
\end{align}
The spread in $r_{W}$ estimates is now $3.1$ times larger than that of $r_{T}$ 
estimates, showing that $\Delta\!\left(t\right)$ does not cause the difference 
in precision, nor does it represent the bias in estimate of $r_{W}$. To complete 
the analysis, we report the sample averages and standard deviations for the 
irreversible system at a sampling time of $20\langle T_{N}\rangle$,
\begin{align}
r_{W} &= 0.2057(7) \pm 0.016 \nonumber \\
r_{T} &= 0.2001(2) \pm 0.004.
\label{eq:estimates_r_irreversible_20}
\end{align}
Now, $r_{T}$ is determined $4.0$ times more precisely than $r_{W}$. We can 
conclude that the estimation error for $r_{W}$ is usually larger than that of 
$r_{T}$ due to a loss of information on the cycle dynamics. For small 
observation times, the estimation errors can actually be similar, but then the 
bias of $r_{W}$ makes it impossible to estimate $r$ correctly. Although the 
exact quantification of the errors and bias remains obscure, we can conclude 
that $r_{T}$ should be preferred as an experimental observable whenever the used 
methods allow for it, as mentioned in Sec.\,IV of the main text. When only 
$r_{W}$ is available due to experimental constraints, the finite-time bias on 
$r_{W}$ should be removed by taking long time traces, or by an appropriate 
fitting procedure that estimates the form of the systematic bias.


%

\end{document}